\documentclass[iop,apjl]{emulateapj}
\usepackage{epsfig}
\usepackage{amsmath, amsthm, amssymb}
\usepackage[plainpages=false, colorlinks=true, anchorcolor=blue, linkcolor=blue, citecolor=blue, bookmarks=false, urlcolor=blue]
{hyperref}
\usepackage{color}
\usepackage{microtype}
\usepackage{float}
\usepackage{natbib}
\usepackage{verbatim}
\usepackage{wrapfig}
\usepackage{graphicx}

\newcommand{\Msun}{$M_\sun$}

\newcommand{\shortauth}{Swihart et al.}
\newcommand{\slugcom}{Draft version - \today}
\slugcomment{\slugcom}

\newcommand{\numcatalinapoints}{471}
\newcommand{\numcatnooutliers}{466}
\newcommand{\mediancatalinamag}{15.72}
\newcommand{\mediancatalinamagerr}{0.06}

\newcommand{\numBpoints}{1643}
\newcommand{\numVpoints}{1038}
\newcommand{\numRpoints}{500}

\lefthead{\sc \footnotesize \slugcom \hfill \shortauth}
\righthead{\sc \footnotesize \slugcom \hfill \shortauth}

\begin{document}

\title{2FGL J0846.0+2820: A new neutron star binary with a giant secondary and variable $\gamma$-ray emission}

\author{Samuel J. Swihart\altaffilmark{1}, Jay Strader\altaffilmark{1}, 
Tyrel J. Johnson\altaffilmark{2},
C.~C.~Cheung\altaffilmark{3},
David Sand\altaffilmark{4},
Laura Chomiuk\altaffilmark{1},
Asher Wasserman\altaffilmark{5},
S{\o}ren Larsen\altaffilmark{6},
Jean P.~Brodie\altaffilmark{5},
Gregory V. Simonian\altaffilmark{7},
Evangelia Tremou\altaffilmark{1},
Laura Shishkovsky\altaffilmark{1},
Daniel E. Reichart\altaffilmark{8},
Joshua Haislip\altaffilmark{8}
}

\affil{ 
  \altaffilmark{1}{Department of Physics and Astronomy, Michigan State University, East Lansing, MI 48824, USA}\\
  \altaffilmark{2}{College of Science, George Mason University, Fairfax, VA 22030, resident at Naval Research Laboratory, Washington, DC 20375, USA}\\
  \altaffilmark{3}{Space Science Division, Naval Research Laboratory, Washington, DC 20375, USA}\\
  \altaffilmark{4}{Department of Physics, Texas Tech University, Box 41051, Lubbock, TX 79409, USA}\\
  \altaffilmark{5}{University of California Observatories, 1156 High Street, Santa Cruz, CA 95064, USA}\\
  \altaffilmark{6}{Department of Astrophysics/IMAPP, Radboud University, P.O. Box 9010, 6500 GL Nijmegen, The Netherlands}\\
  \altaffilmark{7}{Department of Astronomy, The Ohio State University, 140 West 18th Avenue, Columbus, OH 43210, USA}\\
  \altaffilmark{8}{Department of Physics and Astronomy, University of North Carolina at Chapel Hill, Chapel Hill, NC, USA}\\
}

\begin{abstract}
We present optical photometric and spectroscopic observations of the likely stellar counterpart to the unassociated \emph{Fermi}-Large Area Telescope (LAT) $\gamma$-ray source 2FGL J0846.0+2820, selected for study based on positional coincidences of optical variables with unassociated LAT sources. Using optical spectroscopy from the SOAR telescope, we have identified a late-G giant in an eccentric ($e$ = 0.06) 8.133 day orbit with an invisible primary. Modeling the spectroscopy and photometry together lead us to infer a heavy neutron star primary of $\sim 2 M_{\odot}$ and a partially stripped giant secondary of $\sim 0.8 M_{\odot}$. H$\alpha$ emission is observed in some of the spectra, perhaps consistent with the presence of a faint accretion disk. We find the $\gamma$-ray flux of 2FGL J0846.0+2820 dropped substantially in mid-2009, accompanied by an increased variation in the optical brightness, and since then it has not been detected by \emph{Fermi}. The long period and giant secondary are reminiscent of the $\gamma$-ray bright binary 1FGL J1417.7--4407, which hosts a millisecond pulsar apparently in the final stages of the pulsar recycling process. The discovery of 2FGL J0846.0+2820 suggests the identification of a new subclass of millisecond pulsar binaries that are the likely progenitors of typical field millisecond pulsars.

\end{abstract}

\section{Introduction}
\label{sec:intro}
Neutron stars in low-mass binaries can accrete matter and angular momentum from a non-degenerate companion star and be recycled to very fast spin periods, making them detectable as millisecond pulsars \citep[MSPs;][]{MSPs82}. During active accretion, the system is observable as a low-mass X-ray binary. As the orbital period grows and accretion eventually ends, the neutron star turns on as a radio MSP powered by the spindown energy of the neutron star.

Most MSPs in the Galactic field are binaries with a degenerate white dwarf companion of 0.2-0.3~\Msun~and orbital periods in the range of days to weeks \citep{Tauris06}. These are the end products of the recycling process. Recent advances---especially follow-up of sources discovered with the \emph{Fermi} Large Area Telescope (LAT) in GeV $\gamma$-rays---have allowed the discovery of a subclass of MSP binaries with non-degenerate companions in which recycling is apparently not yet complete. These sources, which typically show radio eclipses, are categorized based on their companion's mass. Black widow systems have very light companions ($M_{c}\lesssim0.08M_{\odot}$) that are being actively ablated by the wind of the pulsar, while redbacks have non-degenerate, nearly Roche-lobe filling, main sequence companions of mass $M_{c}\gtrsim0.2M_{\odot}$ \citep{Roberts11}.

Recently, at least three of these redbacks have been found to switch between accretion-powered X-ray binary states and rotationally-powered pulsar states on timescales of days to months \citep{archibald09, Papitto13, Bassa14, Roy15, Bogdanov15, Johnson15}. It is not clear how and when these systems switch on or off as a radio MSPs, nor the cause of the rich phenomenology observed when an accretion disk is present, though both are likely related to the interaction between the inner accretion flow and the pulsar magnetosphere. These transitional MSPs are key systems for understanding the physics of the pulsar recycling process, the interactions between energetic pulsars and their binary companions and surroundings, and the typical evolutionary paths leading to the general MSP population. Current evolutionary models predict such transitions on timescales comparable to the billion-year-long evolution of the binary, not in weeks to months as is observed in the known transitioning systems \citep{Benvenuto15}. Furthermore, these transitional MSPs all have short orbital periods ($\lesssim 0.5$ days); while the evolutionary endpoint of these systems is uncertain, they will likely not be typical of MSP binaries in the field, which have degenerate companions with periods of days to months.

In this work, we report observations of a bright $\gamma$-ray source, 2FGL J0846.0+2820, studied as part of an ongoing survey of unassociated \emph{Fermi}-LAT sources \citep{Strader14, Strader15}. Using follow-up optical observations of the presumed companion, including photometry and spectroscopy, we find that 2FGL J0846.0+2820 is likely a Galactic compact binary with a massive neutron star primary and a giant secondary in a relatively wide orbit. The long orbital period, giant companion, and component masses are remarkably similar to those of the recently discovered $\gamma$-ray bright binary 1FGL J1417.7--4407 \citep{Strader15}, which was independently found to host a radio MSP \citep{Camilo16}. Although no pulsar has yet been discovered to be associated with 2FGL J0846.0+2820, our observations provide evidence of a second system in a new subclass of long-period $\gamma$-ray bright binaries with heavy neutron star primaries and giant secondaries. These systems are plausible progenitors for typical MSP--white dwarf binaries observed in the field.

\section{Observations}
\label{sec:observations}

\subsection{Fermi-LAT Source}
\label{sec:fermiobs}
The $\gamma$-ray source was first detected as 2FGL J0846.0+2820 \citep{Abdo12}, listed in the second full catalog of \emph{Fermi}-LAT sources, based on the first two years of LAT data obtained from 2008 August to 2010 August using the earlier \texttt{P7V6} instrument response functions (IRFs). 2FGL J0846.0+2820 was one of 774 new $\gamma$-ray sources that did not appear in the 1FGL catalog of LAT sources \citep{Abdo10} based on overlapping 95\% source location confidence contours. The overall significance of 2FGL J0846.0+2820 was 4.1$\mathrm{\sigma}$, just above the 2FGL catalog detection threshold of test statistic, TS $>$ 25 (\citealt{Mattox96}, for four total degrees of freedom -- two positional and two spectral; see \S3.2 in \citealt{Abdo12} for a description). With the limited statistics, the average LAT spectrum was best fit as a single power law with photon index $\Gamma$ = 2.51~$\pm$~0.20 and a 0.1--100 GeV flux, $F_{\gamma}$ = (1.1~$\pm$~0.3)~$\times$~10$^{-8}$ photons cm$\mathrm{^{-2}}$ s$\mathrm{^{-1}}$, corresponding to a luminosity of $L_{\gamma}\approx$~(4.0~$\pm$~1.0)~$\times$~10$^{34}$ ($d$ / 8.1 kpc)$^{2}$ erg s$\mathrm{^{-1}}$ (see \S~\ref{sec:distance} for distance estimate), which is comparable to the luminosity of 1FGL J1417.7--4407 \citep[$\sim$3~$\times$~10$^{34}$erg s$\mathrm{^{-1}}$,][]{Strader15} The 2FGL catalog 1-month binned light curve showed only 95\% confidence upper limits indicating each point has TS $<$ 10 or flux error uncertainty $>$ 50\% of the flux value.

We present an updated analysis of the recent Pass 8 \emph{Fermi}-LAT data in \S\ref{sec:fermianaly}.

\subsection{Optical and Near-IR Counterpart}
\label{sec:optical}
The $\gamma$-ray source 2FGL J0846.0+2820 was selected for follow-up study based on a search of positional coincidences of periodic optical variables found in the Catalina Sky Surveys Data Release-1 \citep[CSDR1;][]{Drake14} catalog with \emph{Fermi} LAT \citep{Atwood09} sources. We identified the $V\sim15.7$ Catalina source CSS J084621.9+280839, with a USNO B1.0 catalog \citep{Monet03} J2000 sexagesimal position of (R.A., DEC.) = (08:46:21.89, +28:08:41.0), as a periodic variable 0\fdg215 offset from the 2FGL centroid. This object is also associated with a 2MASS point source with $J = 14.21 \pm 0.02$, $H = 13.59 \pm 0.03$, $K = 13.50 \pm 0.03$ mag \citep{Cutri03}. The CSS source is also listed in the Sloan Digital Sky Survey as J084621.87+280840.8 with magnitudes:  $u = 18.24$, $g = 16.51$, $r = 15.74$, $i = 15.42$, $z = 15.22$ \citep{Ahn12}.

\subsubsection{Catalina Sky Survey (CSS)}
\label{sec:CSS}
To analyze the CSS photometry of this variable, we retrieved~\numcatalinapoints~CSS photometric measurements of the variable optical counterpart taken between 2005 April 10 and 2013 September 25. We searched for periodicity using a Lomb-Scargle periodigram \citep{Scargle82}, finding peak power at a period of $\sim$16.2 d, agreeing with the 16.1866 d period found by \citet{Drake14}. When phased on this period, small periodic flux modulation is evident. However, we have found this period to be an alias of the real orbital period determined via spectroscopy (\S\ref{sec:orbit}). The spectroscopic orbital period is 8.1328 d. Augmented by additional photometry, we analyze the phased light curve of the system in \S\ref{sec:inclination}.

The long-term CSS light curve offers the opportunity to study whether the optical flux from the system has changed significantly since 2005. After removing five measurements that were $>$3$\sigma$ outliers, there are a total of~\numcatnooutliers~measurements. 
The median magnitude is $V_{\textrm{equiv}}$ =~\mediancatalinamag~and the median uncertainty~\mediancatalinamagerr~mag. The long-term CSS 
light curve reveals a monotonic increase in the mean system brightness over the past $\sim$~decade.  A least-square linear regression fit to the data shows the system has brightened at a rate of $0.011\pm0.001$ mag yr$^{-1}$. We discuss this trend as well as the strange phenomenology present in the phased CSS light curves in \S~\ref{sec:longtermbrightness}.

\subsubsection{PROMPT}
\label{sec:PROMPT}

We obtained time series photometry of the optical source in $B$, $V$, and $R$ bands with the 16-inch PROMPT-5 telescope \citep{Reichart05} at Cerro Tololo International Observatory between 2015 February 4 and 2015 June 7. Each observing night consisted of multiple 60-second exposures of the target field, which included the candidate optical counterpart to 2FGL J0846.0+2820 as well as five nearby comparison stars.

We performed differential aperture photometry to obtain instrumental magnitudes of the target source using the five (non-variable) comparison stars as a reference. We calibrated our instrumental magnitudes using observations of the \citet{Landolt92} standard star field RU149. Our final sample includes~\numBpoints~photometric measurements in $B$,~\numVpoints~in $V$, and~\numRpoints~in $R$, with mean magnitudes $B = 16.73$, $V = 15.95$, and $R = 15.40$. All the PROMPT photometry can be found in Table~\ref{tab:PROMPTdata}.

\begin{deluxetable}{cccr}
\tablewidth{170pt}
\tablecaption{PROMPT Photometry of 2FGL J0846.0+2820}
\tablehead{JD-2450000 & Band & Mag & Err \\
                   (d) & & & }

\startdata
7057.65342 & B & 17.001 & 0.289 \\
7057.65421 & B & 16.366 & 0.157 \\
7057.65500 & B & 16.141 & 0.153 \\
7057.65579 & B & 16.683 & 0.214 \\
7057.65658 & B & 16.755 & 0.225 \\
7057.65579 & B & 16.683 & 0.214 \\
7057.65658 & B & 16.755 & 0.225 \\
... & & & \\
\enddata
\tablenotetext{}{Note---This table is published in its entirety in machine-readable format. A portion is shown here for guidance regarding its form and content. These magnitudes are not corrected for extinction.}
\label{tab:PROMPTdata}
\end{deluxetable}

\vspace*{0.5cm}

\section{Optical Spectropscopy}
\label{sec:spectroscopy}

\subsection{SOAR and MDM Spectroscopy}
\label{sec:SOAR}

We began spectroscopic monitoring of the source with the Goodman Spectrograph \citep{Clemens04} on the SOAR 4.1-m telescope on 2014 Dec 8, continuing through 2016 Dec 31. For the initial epochs we used a 2400 l mm$^{-1}$ grating in the region of the Mg$b$ triplet and a 1.03\arcsec\ slit, yielding a resolution of $\sim 0.7$ \AA. The observations from 2015 December onward used a 2100 l mm$^{-1}$ grating, covering a similar region of the spectrum, but with a slightly lower resolution of $\sim 0.9$ \AA. Two or three 600 s exposures were taken per epoch. All spectra were reduced in the standard manner, with wavelength calibration performed using FeAr arcs obtained after each set of spectra. A small number of spectra were obtained with a low-resolution 400 l mm$^{-1}$ grating (resolution $\sim 5.8$ \AA) to check for evidence of emission.

We measured barycentric radial velocities for the medium-resolution spectra through cross-correlation with bright template stars taken with the same setup. Given the long period of the system, at each epoch we determined the radial velocity as a weighted average of the two or three measurements from the individual 600 s exposures. This gave 19 independent SOAR epochs of velocities (with a 20th coming from the Keck/HIRES observation discussed below). These measurements are listed in Table~\ref{tab:RVdata}.

\begin{deluxetable}{crr}
\tablecaption{Barycentric Radial Velocities of J0846}
\tablehead{BJD & RV & Err. \\
                   (d)  & (km s$^{-1}$) & (km s$^{-1}$) }

\startdata
2456999.8170356 & 52.5 & 1.6 \\
2457003.8053888 & 32.5 & 1.6 \\
2457003.8295105 & 33.2 & 1.5 \\
2457012.8609441 & 63.7 & 2.4 \\
2457037.7599617 & 83.0 & 1.7 \\
2457071.6414565 & 97.0 & 1.7 \\
2457119.5191800 & 92.4 & 2.0 \\
2457120.5289206 & 96.3 & 1.7 \\
2457158.4547783 & 36.9 & 1.9 \\
2457166.4751144 & 37.1 & 2.0 \\
2457170.4641598 & 58.6 & 1.9 \\
2457378.8366717 & 66.0 & 2.0 \\
2457417.7571464 & 3.4 & 3.4 \\
2457463.5367120 & 44.1 & 1.9 \\
2457473.6121242 & -12.4 & 2.1 \\
2457495.5359481 & 64.9 & 1.9 \\
2457508.5254554 & 47.7 & 1.9 \\
2457657.1257481\tablenotemark{a} & 95.9 & 0.9 \\
2457744.7835625 & 66.5 & 2.1 \\
2457752.8115122 & 58.4 & 1.8
\enddata
\tablenotetext{a}{This epoch comes from the Keck/HIRES spectrum.}
\label{tab:RVdata}
\end{deluxetable}

In some of our low-resolution spectra we see evidence for H$\alpha$ in emission, though the line is generally weak. To better study the emission, we also obtained some low-resolution spectra with OSMOS on the Hiltner 2.4-m telescope at the MDM Observatory at Kitt Peak.These spectra were obtained in 2--3 exposures of 20 min each on eight epochs from 2016 October 20 to 2017 January 8. Reduced in the usual manner, they cover a usable wavelength range of $\sim 3960$--6840 \AA\ at a resolution of about 3.9 \AA.

These spectra demonstrate a wider range of H$\alpha$ morphology than observed in the small sample of SOAR spectra, from an epoch with very deep H$\alpha$ absorption (1.7 \AA\ equivalent width on 2017 Jan 5) to one with obvious H$\alpha$ emission (about 1.0 \AA\ equivalent width, and substantially larger if corrected for absorption, on 2016 Oct 21). These changes are accompanied by substantial overall changes in the spectral morphology, with a lower effective temperature implied in the Oct 21 spectrum and a higher effective temperature in the Jan 5 spectrum.

\begin{figure}[h]
    \centering
    \includegraphics[width=0.48\textwidth]{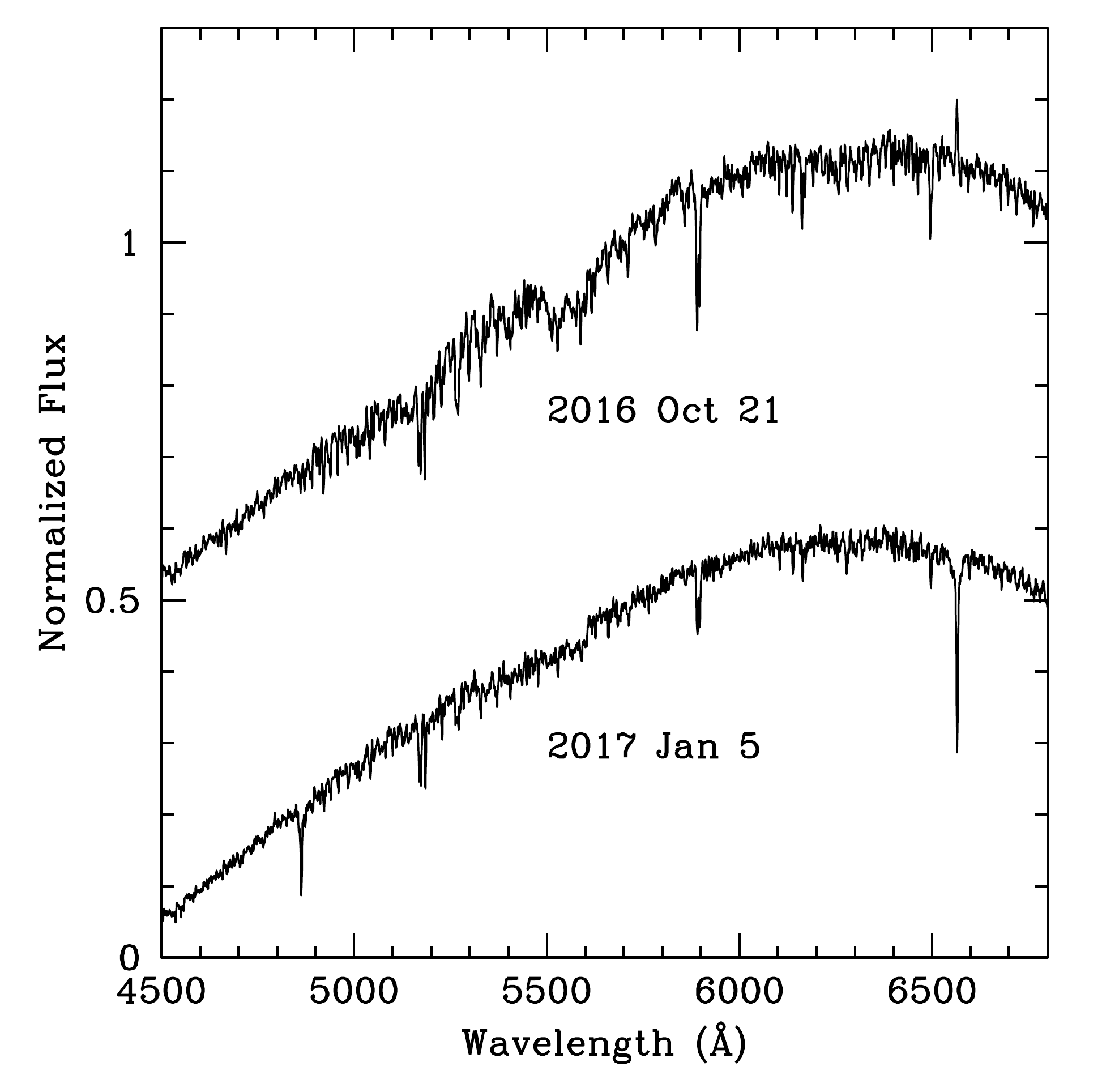}
    \caption{Two of our OSMOS spectra showing the most extreme examples of varying effective temperature and H$\alpha$ emission ($\lambda_{\textrm{rest}}$ = 6562.8~\textrm{{\AA}}). Our other low-resolution spectra lie somewhere between these extremes.}
    \label{fig:tspec}
\end{figure}

In Figure~\ref{fig:tspec} we show a comparison of two OSMOS spectra taken several months apart that illustrate the extremes of varying effective temperature and H$\alpha$ emission. In the 2016 Oct 21 spectrum the effective temperature is lower, with more deeper atomic lines and a hint of the emergence of molecular features, along with clear H$\alpha$ emission and fill-in of H$\beta$. In the 2017 Jan 5 spectrum the effective temperature is clearly higher and there is no longer any H$\alpha$ emission visible; the line appears to entirely be in absorption. Other low-resolution spectra are intermediate between these extremes, though unfortunately we do not have enough spectra to track the temperature or emission effectively as a function of orbital phase.

Analysis of the low-resolution spectra using the spectral classification program {\tt MKCLASS} \citep{Gray14} suggests 
a late-G spectral classification with a subsolar metallicity of --1.1 to --0.6.

\subsection{Keck HIRES Spectroscopy}
\label{sec:vsini}
If the secondary is tidally synchronized with the presumed central compact object, we would expect to see evidence of this through the broadening of spectral lines due to rapid rotation. Using the projected rotational velocity $v$ sin $i$ and orbital semi-amplitude, the mass ratio $q = M_2/M_1$ can be estimated. We found evidence for line broadening in our SOAR spectra, but the resolution was too low to precisely measure $v$ sin $i$.

To address this, we obtained a high-resolution spectrum with HIRES \citep{Vogt94} on the Keck I telescope on 2016 Sep 25. The spectrum was taken using the C5 decker, yielding a nominal resolution of $\sim 36000$, and covered a wavelength range of $\sim$ 3900--8100 \AA. The single 1200-sec exposure was reduced using \texttt{HIRedux} \citep{Bernstein15}.

We additionally obtained a number of spectra of bright late-G to mid-K giant stars with the same resolution and binning to use as templates. We created a set of rotational convolution kernels (assuming a standard limb darkening law) for projected rotational velocities ($v$ sin $i$) 10-100 km $\textrm{s}^{\textrm{-1}}$. Then, on an order-by-order basis, we convolved the spectra of our template stars with the set of kernels and cross-correlated these broadened templates with each of the original, unbroadened spectra. We then fit relations between $v$ sin $i$ and the full-width at half-maximum (FWHM) for each pair of cross-correlations, excluding regions with very strong, broad lines. Finally, we cross-correlated the 2FGL J0846.0+2820 spectrum with the unconvolved template stars and used the FWHM values derived from these cross-correlations to estimate the projected rotational velocity. For $N$ templates, this procedure produces $N^2$ estimates of $v$ sin $i$ per order. In practice, we find that the dispersion in $v$ sin $i$ among templates is much smaller than the dispersion among orders, echoing a similar result found for star cluster velocity dispersions \citep{Strader09}.

The final value derived in this manner is $v$ sin $i$  = 23.2~$\pm$~1.0 km s$^{\rm -1}$, where the uncertainty is the standard deviation of the measurements among all of the templates and orders.

Further examination of the HIRES spectrum showed some evidence of chromospheric activity, and also provided us with an estimate of the foreground reddening from the strength of the NaD lines \citep{Munari97}. The $E(B-V)$ estimate is 0.10, which is slightly higher than the 0.04 predicted from the \citet{Schlafly11} all-sky map.

\section{Results}
\label{sec:results}

\subsection{Keplerian Orbit Fitting and Mass Ratio}
\label{sec:orbit}

\begin{figure}[h]
    \centering
    \includegraphics[width=0.45\textwidth]{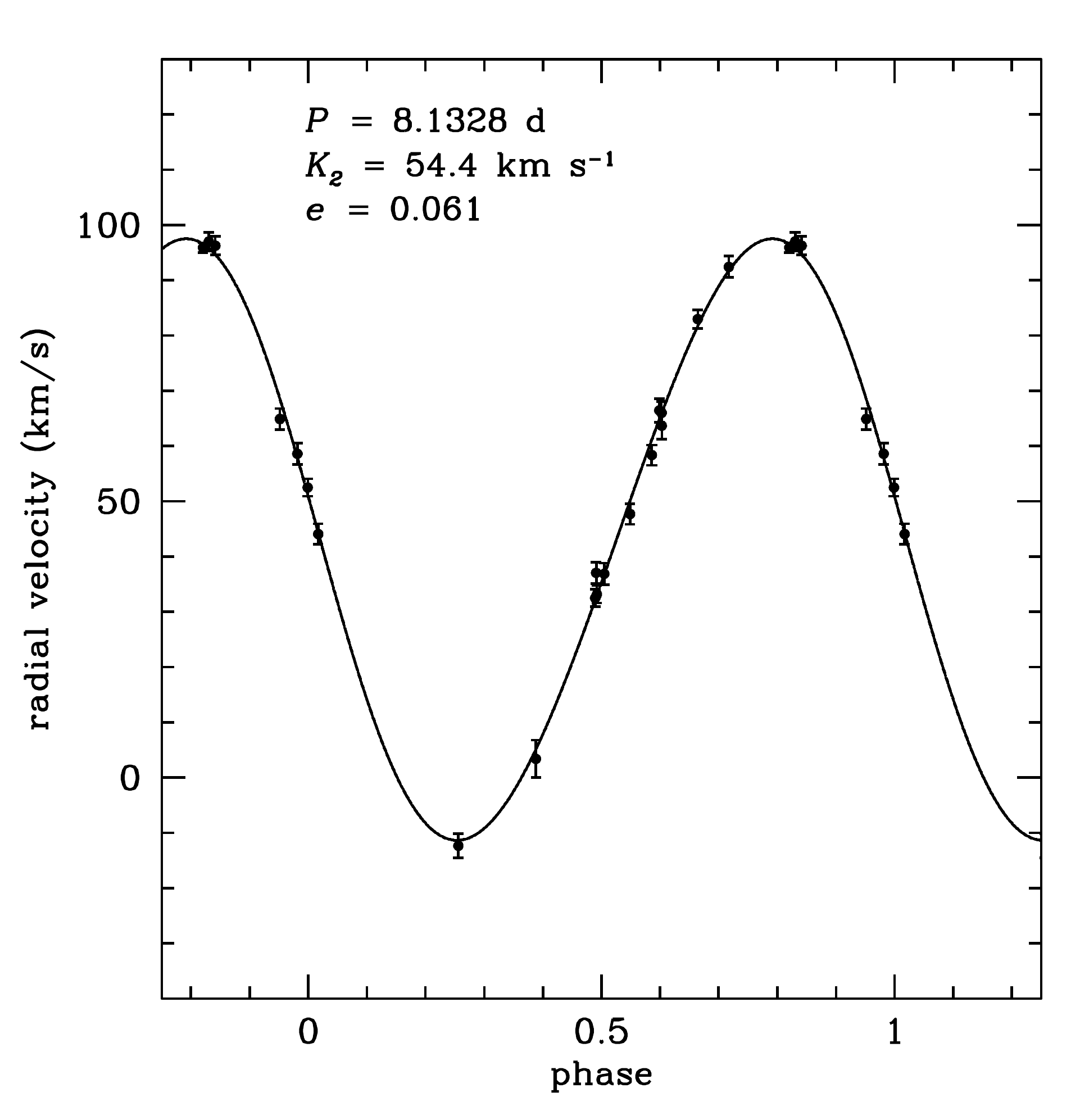}
    \caption{Radial velocity curve for the optical counterpart to 2FGL J0846.0+2820 obtained in 20 epochs between 2014 December 8 and 2016 December 31. The high quality fit yields excellent constraints on the orbital ephemeris.}
    \label{fig:rvfit}
\end{figure}

After correcting the observation epochs to Barycentric Julian Date (BJD) on the Barycentric Dynamical Time system \citep{Eastman10}, we performed a Keplerian fit to our radial velocity data using the custom Markov Chain Monte Carlo sampler \emph{TheJoker} \citep{Price17}. We fit for the period $P$, BJD time of periastron passage $T_{P}$, eccentricity $e$, argument of the periastron $\omega$, systemic velocity $\gamma$, and the semi-amplitude $K_{2}$. The posterior distributions were generally close to normal, with 
equivalent best-fit orbital elements: $P = 8.13284\pm0.00043$ d, $T_P = 2457007.9589_{-0.2907}^{+0.2378}$ d, $e = 0.061\pm0.017$, $\omega = 81\fdg8_{-12.9}^{+10.7}$,
$\gamma = 42.6\pm0.7$ km s$^{-1}$, and $K_2 = 54.4\pm1.0$ km s$^{-1}$. This fit is remarkably good, with an rms for the median values of 1.8 km s$^{-1}$. The phased radial velocity curve is shown in Figure~\ref{fig:rvfit}. 

We use the posterior samples from this fit to derive the mass function $f(M)$

\begin{equation}
f(M) = \frac{P K_{2}^{3}}{2 \pi G} = \frac{M_1 \, (\textrm{sin}\, i)^{3}}{(1+q)^{2}}
\end{equation}

\noindent{for mass ratio $q$ and inclination $i$. We find $f(M) = 0.136_{-0.007}^{+0.008}\,M_{\odot}$. The mass ratio $q$ can be directly determined using our estimate of $v$ sin $i$ and the semi-amplitude of the secondary using the standard formula $v$ sin $i$  = 0.462\,$K_2 \, q^{1/3} \, (1+q)^{2/3}$ \citep{Casares01}. Using the values presented above, this gives $q = 0.402_{-0.037}^{+0.034}$. If the secondary fills its Roche lobe, the orbital period and mass ratio are used to calculate the mean density \citep{Eggleton83}, yielding $\bar{\rho} = 0.003$ g cm$^{\textrm{-3}}$. Together with an analysis of our low-resolution spectra, this result leads us to interpret the secondary as a late-G type giant.}

We defer a detailed discussion of the inclination to \S~\ref{sec:inclination} below, but note that for typical neutron star masses in the range 1.4--2.0 $M_{\odot}$, these measurements suggest an inclination in the approximate range 30--35$^{\circ}$.

\subsection{Fermi-LAT Analysis}
\label{sec:fermianaly}

To examine the $\gamma$-ray source properties with an improved \emph{Fermi}-LAT event reconstruction and instrument characterization, we analyzed $\sim$8.4 years of Pass 8 data from 2008 Aug 04 (15:43:36.0; all UTC times) to 2017 Jan 01 (00:00:00.0), selecting 0.1--100 GeV events within a circular region of interest (ROI) with radius 15$^{\circ}$ centered on the 2FGL position. The photons belonged to the SOURCE class (front and back converted events) as defined under the \texttt{P8R2\_SOURCE\_V6} IRFs and were within a zenith angle of 90$^{\circ}$ of the LAT instrument to minimize contamination from Earth limb photons. We filtered the events to make sure the data were flagged as good and filtered out times corresponding to occurrences of bright, LAT-detected $\gamma$-ray bursts and Solar flares.

Analysis of the $\gamma$-ray data was performed using \emph{Fermi} Science Tools (STs) \texttt{v10r00p05} and binned likelihood analysis. The background model included all 3FGL catalog sources \citep{Acero15} and the diffuse $\gamma$-ray backgrounds \citep{GalModel16} using the respective Galactic and isotropic templates files\footnote{http://fermi.gsfc.nasa.gov/ssc/data/access/lat/BackgroundModels.html}, gll\_iem\_v06.fit and iso\_P8R2SOURCE\_V6\_v06.txt. The quoted uncertainties are statistical only and are larger than expected LAT systematic uncertainties of $\sim$8\% for fluxes and $\sim$0.1 on the spectral slopes \citep{Ackermann12}. The normalization parameters of 3FGL sources with average significance $\geq$ 5$\sigma$ in 4 years and within 6$^{\circ}$ of the ROI center were left free to vary. Additionally, the normalization parameters of sources flagged as significantly variable and within 8$^{\circ}$ of the ROI center, and of the diffuse components were also free to vary.

We first fit the entire $\sim$8.4-year data set centered at the 2FGL J0846.0+2820 position, modeled as a single power law (normalization and index free). The best-fit parameters of this initial fit were $F_{\gamma}$ = (5.2~$\pm$~1.7)~$\times$~10$^{-9}$ photons cm$\mathrm{^{-2}}$ s$\mathrm{^{-1}}$, $\Gamma$ = 3.2~$\pm$~0.4, and point-source TS = 12 ($\sim$3$\sigma$ fitting only two spectral parameters). The flux is significantly less than the 2FGL value (see \S~\ref{sec:fermiobs}), implying a variable source. To quantify the variability, we constructed a $\sim0.5$-year binned light curve with 17 bins, the first 16 of which had a 180 day duration and the last one 191 days. For the fit in each time bin, we allowed the same parameters to vary as in the fit of the entire data set but kept the photon index of 2FGL J0846.0+2820 fixed to the initially preliminary fit value. The $\gamma$-ray emission was found to be concentrated within the first two 180-day bins with TS = 6.2 and 29.5 respectively, coinciding with the first half of the time interval analyzed in the 2FGL catalog.

As a next step, we checked for any nearby point sources new to the 8.4-year dataset and not in the 3FGL 4-year catalog. This was done by generating a large TS map of the field, spanning 5$^{\circ}$ $\times$ 5$^{\circ}$, with $0\fdg25$ per pixel. We found two candidate sources, and used \texttt{gtfindsrc} to localize them, then refit their spectral parameters with \texttt{gtlike} adopting single power-law models. Only one of the candidates appeared significant and was included in the model, with $F_{\gamma}$ = (4.1~$\pm$~1.3)~$\times$~10$^{-9}$ photons cm$\mathrm{^{-2}}$ s$\mathrm{^{-1}}$, $\Gamma$ = 2.3~$\pm$~0.4, and TS = 43.0. The $z=1.283$ blazar, B2 0849+28 \citep{Hewitt10}, is just $0\farcm7$ offset from its best-fit LAT position, R.A. = 133\fdg028, Decl. = +28\fdg556, and $95\%$ confidence error radius, $r_{\rm 95} = 4\farcm6$, and is likely the source of the $\gamma$-ray emission.

\begin{figure}[h]
    \centering
    \includegraphics[width=0.50\textwidth]{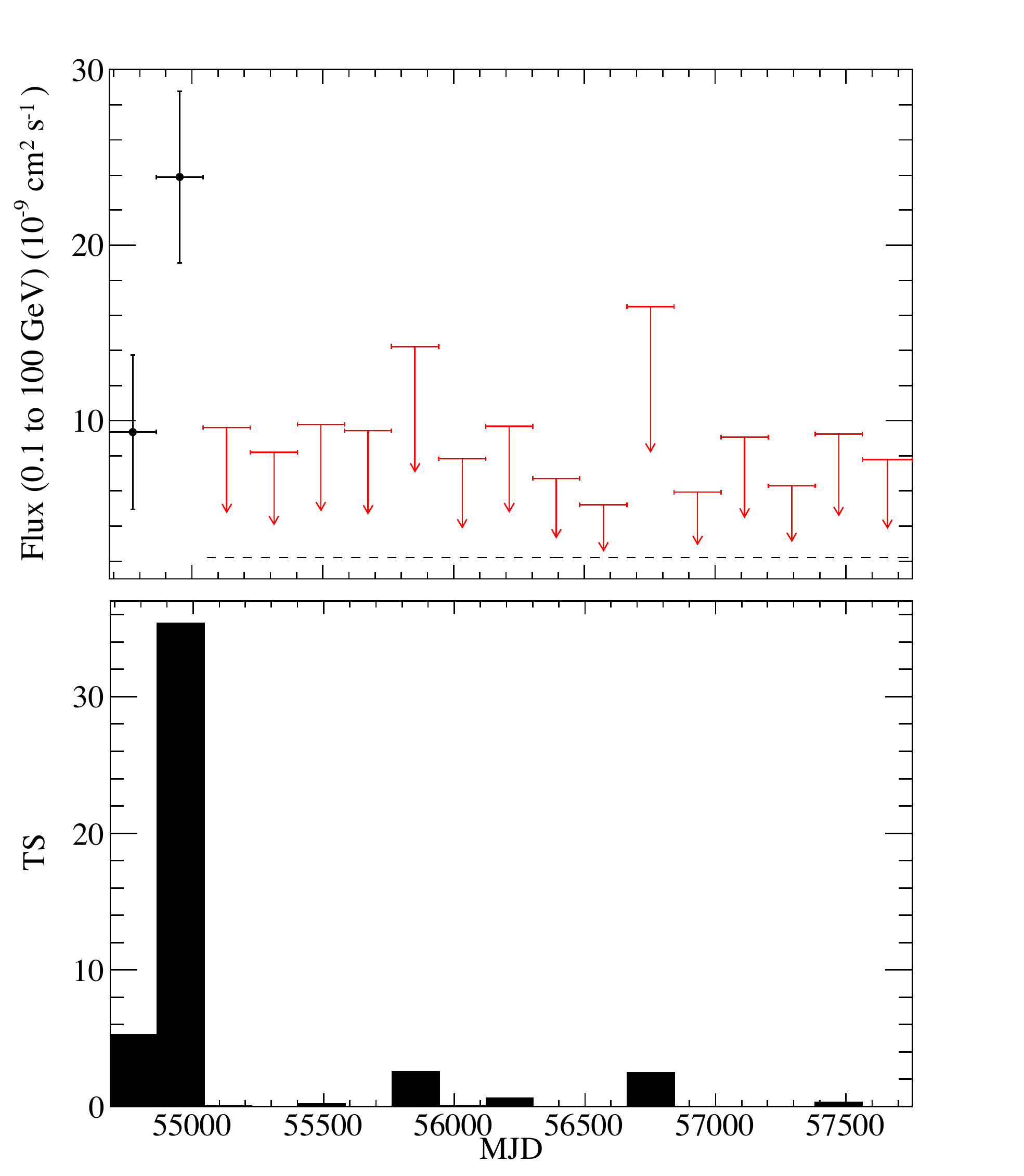}
    \caption{\emph{Fermi}-LAT 0.1--100 GeV $\gamma$-ray light curve (Top) and corresponding TS values (Bottom) in $\sim$~0.5 year bins. In the light curve, down arrows indicate 95\% confidence upper limits on the fluxes, and the upper limit derived from the data after the first two bins is indicated with the horizontal dashed line.}
    \label{fig:fermiflux}
\end{figure}

With the new point source in the background model, we fit the LAT position of the target source with only the first two 180-day bins of data, resulting in R.A = 131\fdg83, Decl. = 28\fdg17 (0\fdg212 from the CSS source), $r_{\rm 95} = 0\fdg14$ and average flux, $F_{\gamma}$ = (2.2~$\pm$~0.7)~$\times$~10$^{-8}$ photons cm$\mathrm{^{-2}}$ s$\mathrm{^{-1}}$ ($\sim$2x the flux measured when fitting the entire dataset (\S~\ref{sec:fermiobs})), $\Gamma$ = 2.7~$\pm$~0.2, and TS = 35. For the same 360-day dataset, we compared the best-fit likelihood for a exponentially cutoff power law (normalization, index, and cutoff energy free) and found no significant evidence for curvature in the spectrum \citep[TS$_{\rm cut}$ $= 0.0$; see][]{AbdoAA13}. For completeness, we also repeated the analysis with the full 8.4-year dataset, and confirmed the initial finding of a low-significance source (TS = 6.4) averaged over the larger dataset. We utilized the fitted index = 2.7 that was  based on the first 360-days of data, and the new best-fit LAT position to generate the final $\sim0.5$-year lightcurves as presented in Figure~\ref{fig:fermiflux} where we only report points for bins in which 2FGL J0846.0+2820 was found with both TS~$\geq$~4 ($\sim$2$\sigma$) and $\geq$~4 predicted counts, else a 95\% confidence upper limit is shown. We confirmed the initial result that the $\gamma$-ray emission is concentrated within the first two 180-day bins with TS = 5.3 and 35.4, respectively. We quantify the significance of variability as TS$_{\textrm{var}}$ = 161.6 following the \emph{Fermi}-LAT catalog analysis \citep[see \S3.6 in][for the definition]{Abdo12}, which is well above the threshold of 32 needed to flag this source as variable at 99\% confidence level.

\begin{figure}[h]
    \centering
    \includegraphics[width=0.45\textwidth]{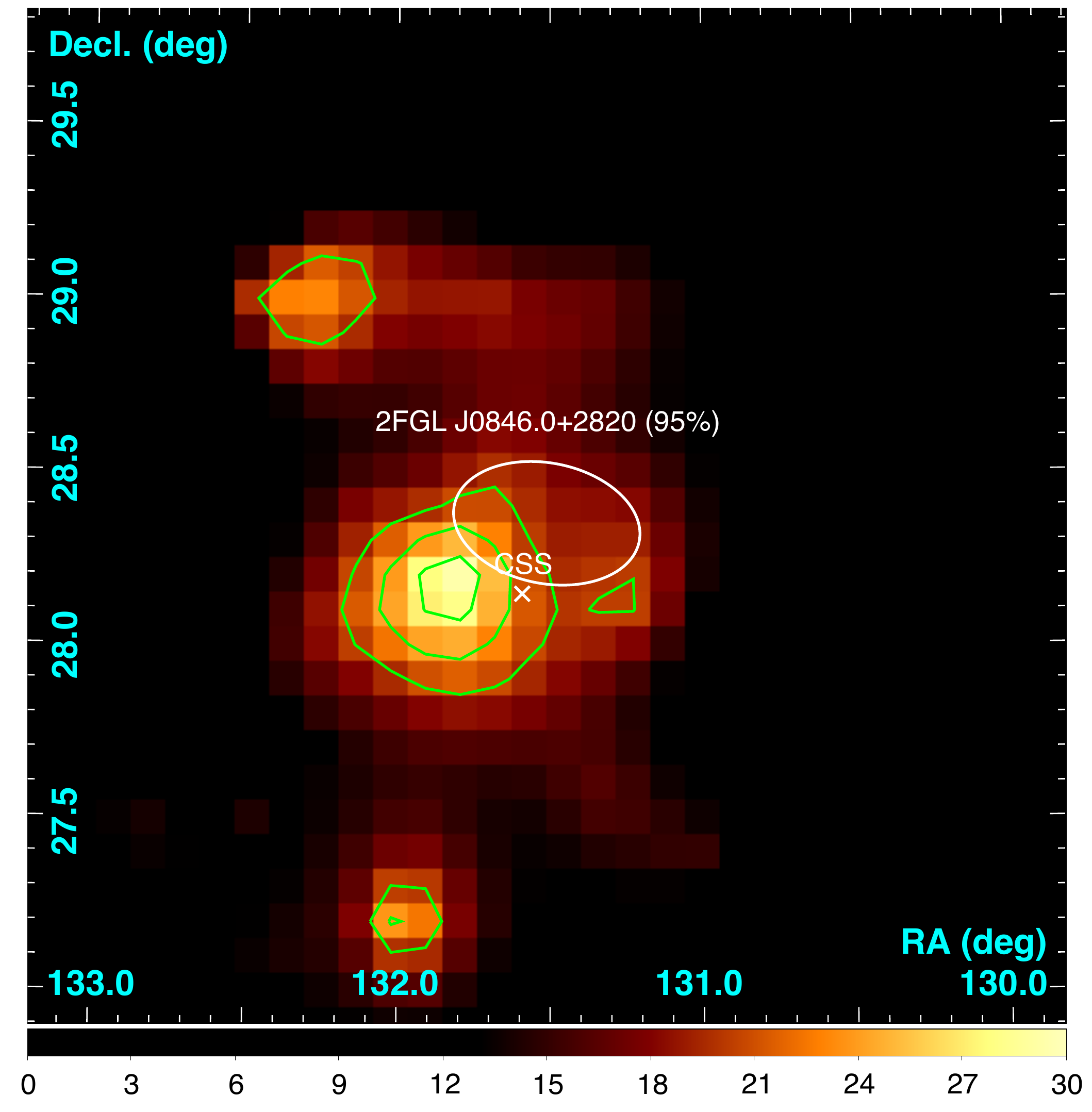}
    \caption{The LAT TS map centered on 2FGL J0846.0+2820 in J2000.0 coordinates (bottom bar indicates TS values). The green contours are, from inside out, 68, 95, 99\% confidence on the localization, newly determined from the Pass 8 data analysis. The 2FGL 95\% confidence ellipse (white ellipse) is shown as well as the optical CSS position (white cross).}
    \label{fig:TSmap}
\end{figure}

We produced a TS map of the region surrounding 2FGL J0846.0+2820 in a grid of 0\fdg1 per pixel side using the \emph{Fermi} ST \texttt{gttsmap} and data from the first 360 days of the \emph{Fermi} mission, and a model which did not include a source corresponding to 2FGL J0846.0+2820. The resulting TS map is shown in Figure~\ref{fig:TSmap} with the localization contours, 2FGL ellipse, and optical position. The optical position is just outside both 95\% contours, but within the new 99\% localization. The new Pass 8 localization and 2FGL 95\% contours overlap indicating these are the same source, however, the shift in the position is notable despite this being located at a fairly high Galactic latitude, $b = 36\fdg3$, and our work in identifying all possible significant point sources in the field using the 8.4-year dataset. The shift could be due to several factors, including using different event reconstructions (P7V6 IRF) and time range compared to the 2FGL catalog analysis, diffuse models for the diffuse background models, and could also indicate a region with poorly modeled diffuse emission seen as extended regions of signal in the residual TS map. To investigate the regions of excess signal visible in Figure~\ref{fig:TSmap} that are not associated with 2FGL J0846.0+2820, we constructed another TS map with our target source in the model at the new position. This resulted in no excesses above TS = 13.5 and confidence contours which indicated possible extension, suggesting a positive fluctuation of unmodeled diffuse emission (there are no known features in the adopted diffuse emission model in the region of this source; see Fig. 5 in \citealt{GalModel16}). This is further supported by the fact that no significant TS excesses were observed at these positions in the TS map covering the full time range.

\begin{figure}[h]
    \centering
    \includegraphics[width=0.46\textwidth]{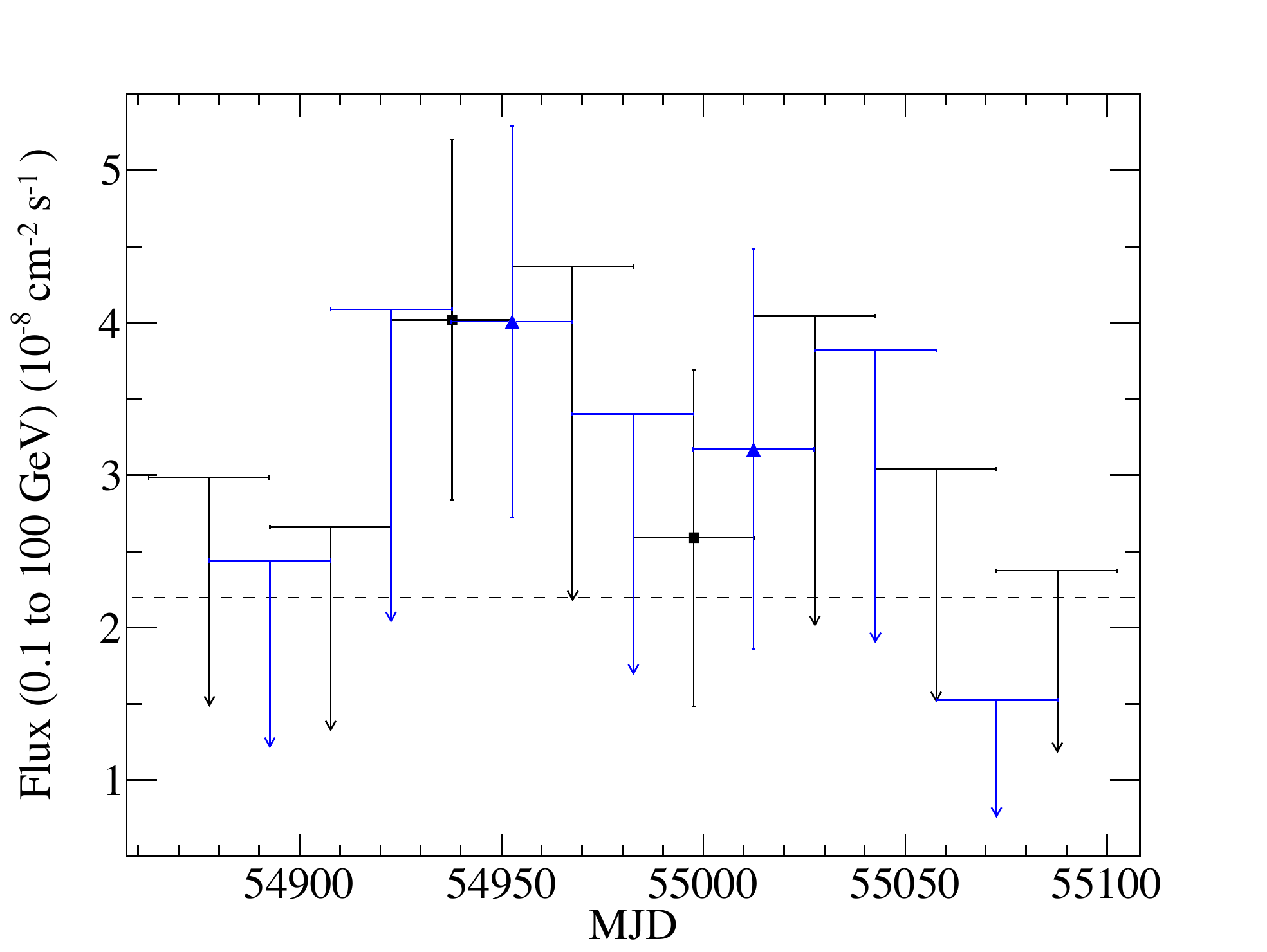}
    \caption{The LAT light curve for the most significant 180-day interval ($\pm$60 days) in shorter 30-day time bins (black points). The data were shifted by 15 days (blue points) to find the last significant time bin. The upper limits are indicated when TS $<$ 4 with fewer than four predicted photons. The dashed line marks the best fit average flux.}
    \label{fig:fluxshifted}
\end{figure}

In the interest of better constraining when the flux of 2FGL J0846.0+2820 dropped, we analyzed the second 180 day time bin (2009 Jan 31 to 2009 July 31) in more detail. We constructed flux light curves with 30-day bins spanning $\pm$60 days of our 180 day time bin. We then constructed a second 30-day flux light curve with the start times shifted by 15 days (Figure~\ref{fig:fluxshifted}). Due to the rather faint nature of this source, we can not break the flux light curve into smaller intervals, but we estimate that the flux dropped sometime between 2009 May 31.7 (MJD 54982.7) and 2009 Jul 30.7 (MJD 55042.7).

An analysis fitting the entire LAT data set except the first 360 days, with the photon index fixed ($\Gamma = 2.7$), resulted in a non-detection with TS = 0 and a 95\%  flux upper limit of $<$2.2~$\times$~10$^{-9}$ photons cm$\mathrm{^{-2}}$ s$\mathrm{^{-1}}$. This amounts to a drop in flux of $\sim$10x with respect to the second 180-day detection.

Using the spectroscopic optical period and the second (and most significant) 180 day time bin, we searched the LAT data for evidence of modulation with the orbit. In order to do so, we calculated spectral weights, calculated with the best-fit model for this time range and the \emph{Fermi} ST \texttt{gtsrcprob}, which have been shown to enhance sensitivity to periodic signals \citep[e.g.,][]{Kerr11}. We then calculated the exposure at the position of 2FGL J0846.0+2820 in 30 second intervals over this time period in order to correct for exposure differences with orbital phase \citep[as described in, for instance,][]{Johnson15}. We used the \emph{Fermi} ST \texttt{gtpphase} to assign orbital phase values to the LAT events and, correcting for exposure and using the spectral weights, tested for modulation using the H-test \citep{Jager89, Jager10} and the Z$^{2}_{\mathrm{m}}$ test with two harmonics. Both tests showed no significant modulation, returning 0.1$\sigma$ and 0.3$\sigma$, respectively.

2FGL J0846.0+2820 is one of 234 unassociated sources not present in the 3FGL catalog, despite a detection at high significance in the earlier (2FGL) catalog. Many of the sources lost between 2FGL and 3FGL were at low Galactic latitudes where the Galactic diffuse emission is strongest, such that improvements in modelling this emission was expected to have the most influence in detecting sources. However, 2FGL J0846.0+2820 is out of the Galactic Plane. A number of 2FGL sources out of the Plane were also spurious, some apparently due to contamination from the Moon \citep{Corbet13}. However, at the declination of 2FGL J0846.0+2820 (ecliptic coordinates: ($\lambda$, $\beta$) = (126\fdg4, 9\fdg8)), the effects of the Moon should be minimal, and given the refined event reconstruction and characterization presented here, there is no compelling evidence that it was a spurious source. Instead, our analysis shows that the most likely explanation for the absence of the source from the 3FGL catalog is the source variability over the lifetime of \emph{Fermi}.

\subsection{Optical Light Curve Models}
\label{sec:inclination}
\begin{figure}[htbp]
    \includegraphics[width=0.48\textwidth]{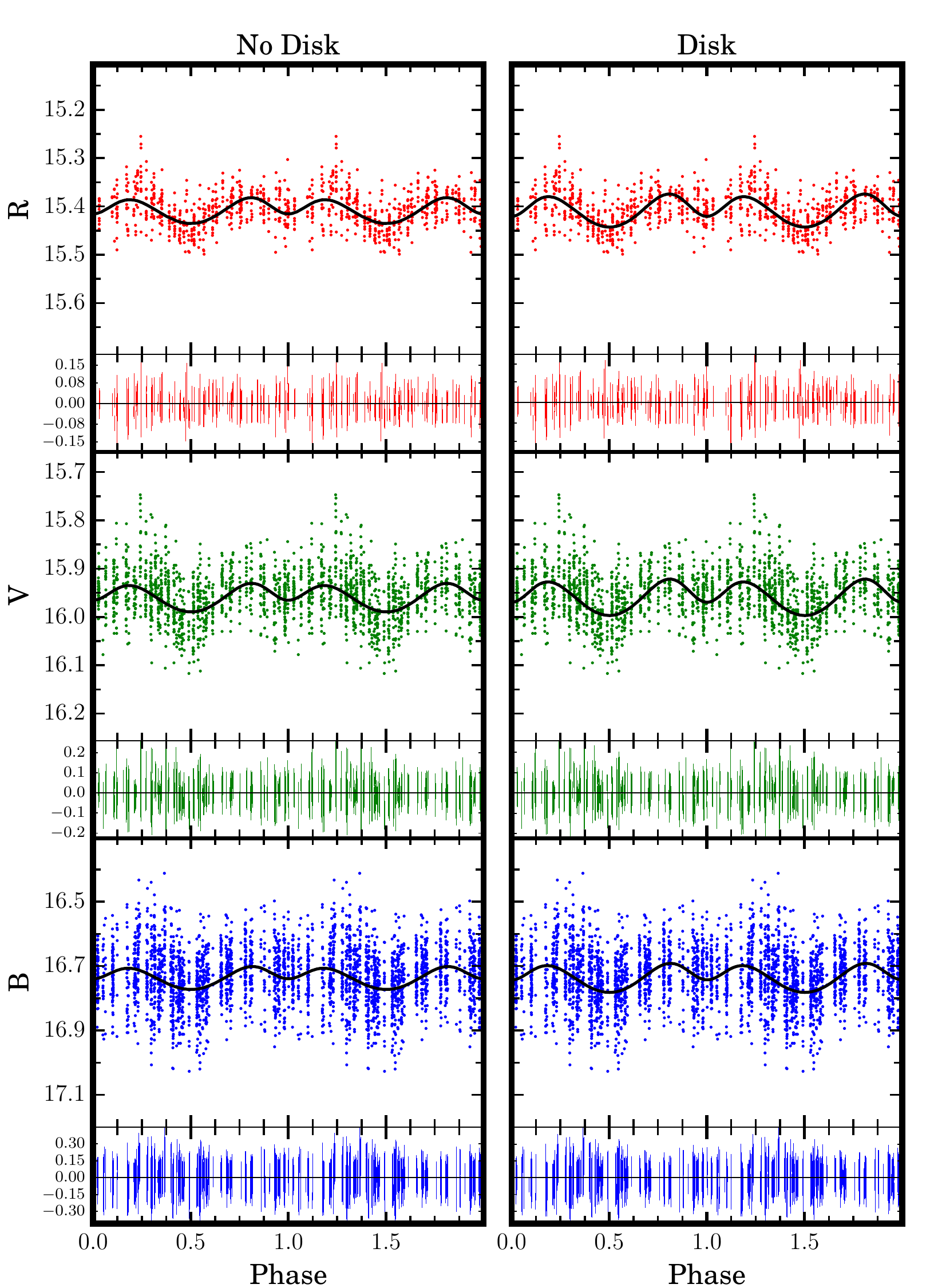}
    \centering
    \caption{PROMPT $B$, $V$, and $R$ light curves folded on the spectroscopic period. The best fit ELC models that included (right) and excluded (left) a steady accretion disk component are shown with black lines. Uncertainties for each measurement are not shown in the upper panels, but are displayed in the fit residuals (lower panels). Two orbital phases are shown for clarity.}
    \label{fig:ellipsoidal}
\end{figure}

We show the phased light curves of the binary in Figure~\ref{fig:ellipsoidal}. There are two maxima and minima per orbit, consistent with ellipsoidal variations due to a tidally deformed secondary orbiting the presumed neutron star primary. Motivated by the H$\alpha$ emission and change in the $\gamma$-ray flux of the binary, we start from the assumption that the secondary is filling its Roche lobe, and then explore models where the Roche lobe filling factor is free to vary.

The baseline expectation for ellipsoidal variations are two equal maxima when the projected area of the tidally distorted star is largest, and two unequal minima due to varying effects of gravity darkening when the system is viewed along the axis connecting the primary and secondary. Modelling these ellipsoidal modulations can constrain the inclination of the system between the extremes where the maximum effect is observed if the system is edge-on ($i = 90^{\circ}$) and no modulations are expected for face-on inclinations ($i = 0^{\circ}$).

The broad filter used by CSS and its relatively large photometric uncertainties do not make these data ideal for modeling ellipsoidal variations in 2FGL J0846.0+2820. Instead, to model the light curve we use the more precise and well-sampled PROMPT photometry in $B$, $V$, and $R$ bands.

We model these light curves using the Eclipsing Light Curve \citep[ELC;][]{ELCcode} code. We try fits that include contributions from a Roche-lobe filling secondary and a steady accretion disk component, as well as fits that exclude the disk contribution and allow the filling factor to vary. Throughout, we assume the primary object is invisible. In addition, we set the scale of the system by fixing the value of the orbital period, semi-amplitude, binary mass function, mass ratio, and eccentricity to those determined via spectroscopy. These values imply an average orbital separation of $a \sim 26\,R_{\odot}$.  We also require the observed value of $v$ sin $i$ be matched.
From an analysis of our low-resolution spectra and the photometric colors, we fix the effective temperature of the secondary ($T_2$) to 5250 K. We note this value is not well-constrained as the ELC code fits normalized light curves in all bands and is thus insensitive to the effective temperature of the companion. The assumed effective temperature does affect the inferred distance, an issue we revisit in the following section. Given the long period of the system and its luminous secondary, irradiation of the companion by a central X-ray source is likely to be less important than in typical black widows or redbacks. While \emph{Chandra} X-ray observations of 2FGL J0846.0+2820 are forthcoming, we have only upper limits on its X-ray flux from all-sky monitors (no clear detections from the \emph{Monitor of All Sky X-ray Image} \citep[MAXI, $\sim$1-20 keV;][]{Matsuoka09} or the Swift \emph{Burst Alert Telescope} \citep[BAT, $\sim$15-150 keV;][]{Barthelmy05}), which is a comparably weak constraint. For an X-ray luminosity of 10$^\textrm{{32}}$-10$^\textrm{{33}}$ erg s$\mathrm{^{-1}}$, the ratio of optical to X-ray flux at the surface of the star is $>$100, suggesting irradiation is not very important.
On the other hand, recent detailed modeling of intra-binary shocks in redbacks have suggested the pulsar wind flow is channeled onto the stellar surface due to the magnetic field of the secondary \citep{Romani16, Wadiasingh17}, which may explain the unmodeled features seen in the light curve, such as a phase shift, in the context of irradiation. This issue should be explored in more depth when future X-ray data are available.
 
We fit the folded light curves for $B$, $V$, and $R$ bands simultaneously. 
For models that included an accretion disk component, we fixed the filling factor of the secondary to 1 while fitting the following parameters: the inclination $i$, the inner disk temperature $T_{\textrm{disk}}$, the opening angle of the disk rim $\beta$, the inner ($r_{\textrm{in}}$) and outer ($r_{\textrm{out}}$) radii of the disk, the power-law index of the disk temperature profile $\xi$, and a small phase shift $\Delta \phi$. For models that did not include a disk, we fit the inclination, the filling factor of the secondary $f_2$, and a phase shift. The best fit parameters with 1$\sigma$ confidence levels are listed in Table~\ref{table:bestfits}. Parameters without listed errors were not well constrained by the models (i.e. flat $\chi^2$ distributions) so we refrain from quoting their exact uncertainties, and instead only report the values associated with the best fit.

\begin{deluxetable}{lcr}[htbp]
\tablecaption{Best fit ELC parameters}
\tablehead{Free Parameter & No Disk & Disk}

\startdata
R.L. filling factor\tablenotemark{a} & 0.86 $\pm$ 0.03 & 1.0 \\[2pt]
Inclination ($^{\circ}$) & {27.1}$^{+1.1}_{-1.0}$ & {30.7}$^{+3.0}_{-1.7}$ \\[2pt]
Inner disk temp. (K) & --- & 3600$^{+4600}_{-2500}$ \\[2pt]
Disk opening angle ($^{\circ}$) & --- & 15.0 \\[2pt]
Inner disk radius\tablenotemark{b} & --- & 0.05 $\pm$ 0.04 \\[2pt]
Outer disk radius\tablenotemark{b} & --- & 0.70 $\pm$ 0.13 \\[2pt]
Disk powerlaw index ($\xi$) & --- & --0.62 $\pm$ 0.2 \\[2pt]
Phase shift ($\Delta \phi$) & --0.022 $\pm$ 0.006 & --0.024 $\pm$ 0.005 \\[2pt]
$\chi^2$(dof) & 3987.0 (3178) & 3989.6 (3174)

\enddata
\tablenotetext{a}{The Roche-lobe filling factor of the secondary was fixed to 1.0 for all models that included a disk.}
\tablenotetext{b}{Expressed as a fraction of the effective Roche-lobe radius of the primary.}
\label{table:bestfits}
\end{deluxetable}

The first clear result from the fits is that a small phase shift ($\Delta \phi$) is required: models with no shift can be clearly ruled out at the 4--5$\sigma$ level. Approximately the same $\Delta \phi$ was found for both the disk and no-disk case.
Due to the long period, the small value of $\Delta \phi$ actually corresponds to a substantial offset of $\sim 4.3$--4.7 hr, depending on the exact model.

Similar phase offsets have been observed in some black widow and redback systems, likely due to asymmetries in the heating of the secondary or in the disk \citep[e.g.,][]{Romani16, Li16}. One possibility is that these asymmetries may be caused by hot spots in a disk or star spots on the companion. Although the ELC code allows us to fit for a variety of disk or star spots, we find that adding such features do not significantly change the fit quality, probably due to the relatively face-on nature of the system and our typical measurement uncertainties. Nonetheless, starspots may be important for understanding the long-term variability in the system (see \S~\ref{sec:longtermbrightness}).

For models that include a disk component and a Roche-lobe filling secondary, we find our best fits to the data occur when the accretion disk contributes a small fraction ($<$3\%) of the total light from the system. This relatively low veiling is not necessarily unexpected, given the weak H$\alpha$ emission and the dominance of the photometric light curve by ellipsoidal variations. For fits with a disk we find an inclination of 30\fdg7{$^{+3.0}_{-1.7}$}.

For models without a disk,the best fit $\chi^2$ is slightly lower than our best ``with disk'' model, but given the large number of data points the difference is negligible. Our best fit occurs when the companion is slightly underfilling its Roche lobe, with a filling factor $f_2=0.86$. This is very similar to the results of \citet{McConnell15} for the transitional MSP J1023+0023, who found  $f_2 = 0.83^{+0.03}_{-0.02}$ while the system was in the radio MSP state.

For 2FGL J0846.0+2820, the inclination implied by the no-disk fit is 27\fdg1{$^{+1.1}_{-1.0}$}. This value is slightly lower than inferred for the fit including a disk, consistent with the general result that disk veiling leads to an underestimate of the inclination from ellipsoidal variations \citep[e.g.,][]{Kreidberg12, McConnell15, Wu15}.

\subsubsection{Distance}
\label{sec:distance}
We estimate the distance to the system by comparing the luminosity inferred from the best-fit secondary radius and assumed temperature to the observed magnitude of the system. We determined bolometric corrections using 10 Gyr isochrones assuming [Fe/H] = --1 \citep{Marigo08}. The best fit model assuming the star fills its Roche lobe has an effective radius of $7.1\,R_{\odot}$. Using an assumed effective temperature of 5250K, the secondary has a predicted $M_V = 1.1$ mag. Compared to the observed $V_0 = 15.63$ mag (corrected for extinction using $E(B-V) = 0.10$), the distance is $\sim 8.1$ kpc. For an effective temperature of 5000 K, the 
distance would instead be $\sim 7.0$ kpc.

We performed similar calculations for the case of an underfilled Roche lobe, finding very similar results: $\sim$ 8.3 and 7.2 kpc for temperatures of 5250 K and 5000 K, respectively. We emphasize that these distance calculations are uncertain and dominated by systematic effects.

Even given the large uncertainty in the distance, the location of the system in the Galactic anti-center ($l = 197^{\circ}$) and its substantial distance above the plane ($\sim 4$ kpc for $b = 36^{\circ}$) makes it unlikely that the binary was formed in the thin disk.

\subsection{Component Masses}
\label{sec:masses}

Using the posterior samples for the orbital period, semi-amplitude, and mass ratio obtained from our spectroscopic analysis (\S~\ref{sec:orbit}), along with our estimates of the inclination from fitting the photometry, we can infer the primary and secondary masses of the system. For the fits that excluded a disk, we find $M_{1} = 2.81 \pm 0.36\,M_{\odot}$ and $M_{2} = 1.12 \pm 0.21\,M_{\odot}$. The primary mass implied by these models is larger than any known neutron star and approaches 
the maximum theoretical mass \citep{Chamel13}.

For fits that included a disk, our models imply masses $M_{1} = 1.96 \pm 0.41\,M_{\odot}$ and $M_{2} = 0.77 \pm 0.20\,M_{\odot}$.
In contrast to the fits with no disk, the primary mass is more typical of those inferred for neutron stars in binaries 
associated with \emph{Fermi} sources \citep[e.g.,][]{Ransom11, Romani15, Strader15, Strader16}, and we think these masses are more likely to be accurate.

The uncertainties in the component masses are large due to the relatively face-on inclination. Furthermore, the potential for indirect, asymmetric heating on the face of the secondary may be adding an additional source of systematic uncertainty to our mass estimates. A few previous publications that describe the light curve modeling of similar black widow and redback systems have inferred unusually large neutron star masses ($>$2\Msun), which were later found to be unreliable because of inadequacies in the assumed model \citep[e.g. not accounting for indirect heating via pulsar spin-down power reprocessed in an intra-binary shock, see][and references therein]{Romani15_2, Romani15}. However, given the clear differences between this system and typical MSP spiders, including the luminous secondary, the wide orbit, and the face-on orientation of the binary, unmodeled irradiation is expected to make a less important contribution to the systematic uncertainties in our mass estimates than for most MSP binaries with hydrogen-rich companions (see discussion in \S~\ref{sec:inclination}).

Nonetheless, other interpretations of the system are unlikely: neither stellar-mass black holes nor white dwarfs in quiescence have been observed to emit GeV gamma rays. The inclination would need to be considerably lower ($i = 21^{\circ}$) to be consistent with even a low-mass $5\,M_{\odot}$ black hole, and would need to be somewhat higher (35--40$^{\circ}$) to accommodate a white dwarf. Furthermore, an accreting white dwarf with a giant secondary in an 8.1 d orbit would be extraordinarily unusual; most such ``symbiotic" systems have periods of hundreds of days and accrete from a wind \citep{Belczynski00}. Future observations can help improve constraints on the presence of a disk and hence on the inclination.

\subsection{Long-Term Optical Brightness}
\label{sec:longtermbrightness}
\begin{figure}[h]
    \includegraphics[width=0.48\textwidth]{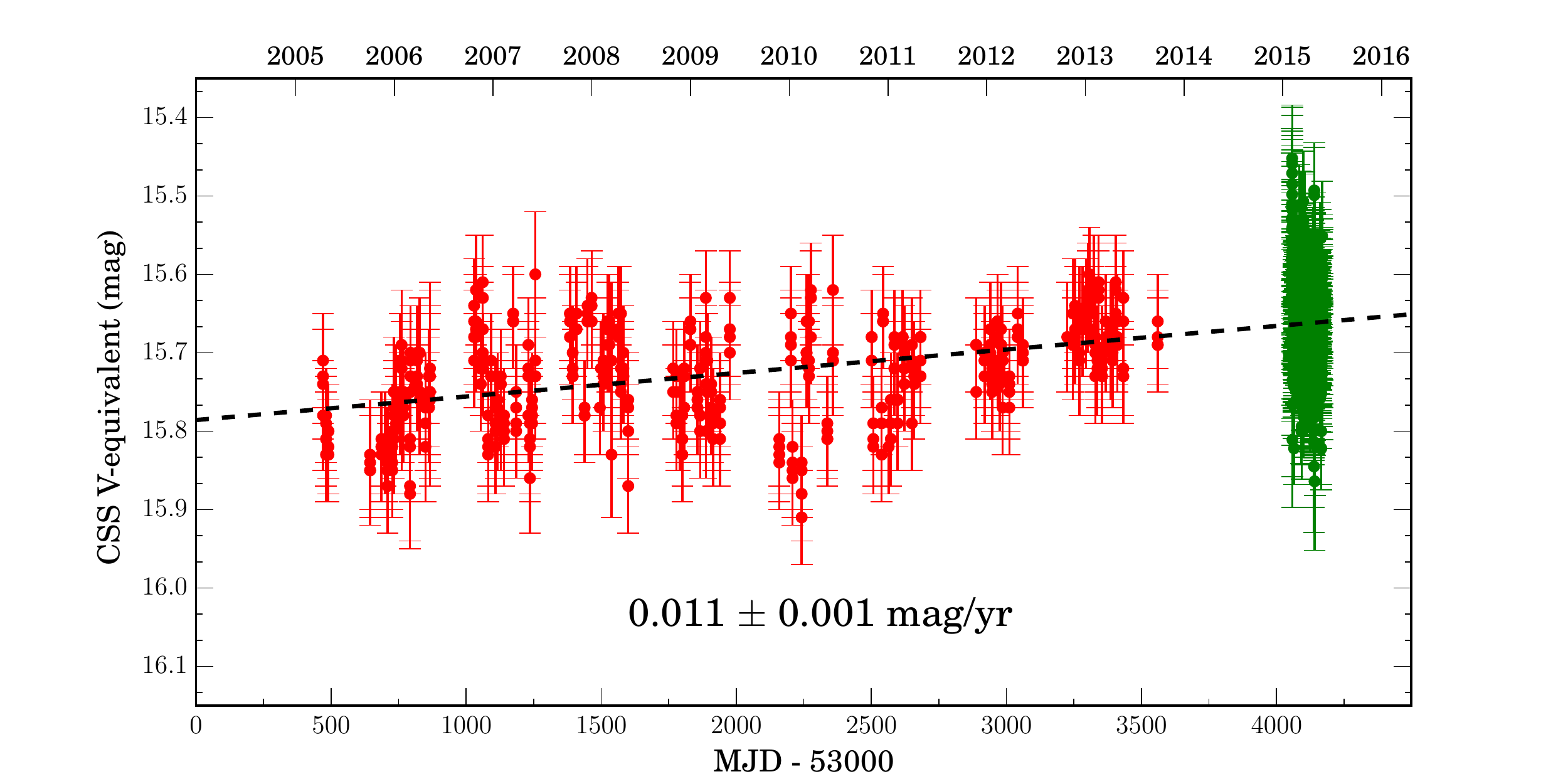}
    \centering
    \caption{Long-term optical light curve showing a monotonic increase in the system brightness over the past decade. The Catalina data are shown in red and our recent PROMPT $V$-band data are in green. The dashed line and rate of brightening are the result of a linear fit to the Catalina data only. A small zeropoint offset has been applied to the PROMPT data as described in \S~\ref{sec:longtermbrightness}}
    \label{fig:catalina}
\end{figure}

As discussed above, there is compelling evidence that the system has brightened at a rate of $0.011\pm0.001$ mag yr$^{-1}$ over the last decade, with only a few small interruptions in this monotonic trend (Figure~\ref{fig:catalina}). There are a number of possible explanations for this trend: the secondary could be slowly increasing in radius; the mean effective temperature could be increasing (either globally or due to a change in the properties of starspots); or an accretion disk could be getting brighter. The last of these was proposed as an explanation for similar secular brightness trends in the quiescent black hole binary systems Nova Muscae 1991 \citep{Wu15} and A0620-00 \citep{Bailyn16}. 

The trend for 2FGL J0846.0+2820 is not confined to the CSS data: we find that our recent PROMPT data fits the trend, after correcting for a small zeropoint offset of -0.30 mag in the listed Catalina magnitudes using non-variable comparison stars in common between the PROMPT and CSS data.

In addition to the overall trend, the data taken between 2009 November and 2010 June shows a larger scatter than observations before or after this.
Above (\S\ref{sec:fermianaly}), we showed the $\gamma$-ray emission disappeared around the start of 2009 July, and these high-scatter observations are the first CSS data taken after this change in the $\gamma$-ray flux. Given the long time baseline for comparison, the photometric behavior around this epoch is unusual, and we consider it unlikely that these two events are coincidental: 
\emph{something} happened in mid-2009 that affected both the $\gamma$-ray flux and optical brightness of the binary.

\begin{figure}[h]
    \includegraphics[width=0.48\textwidth]{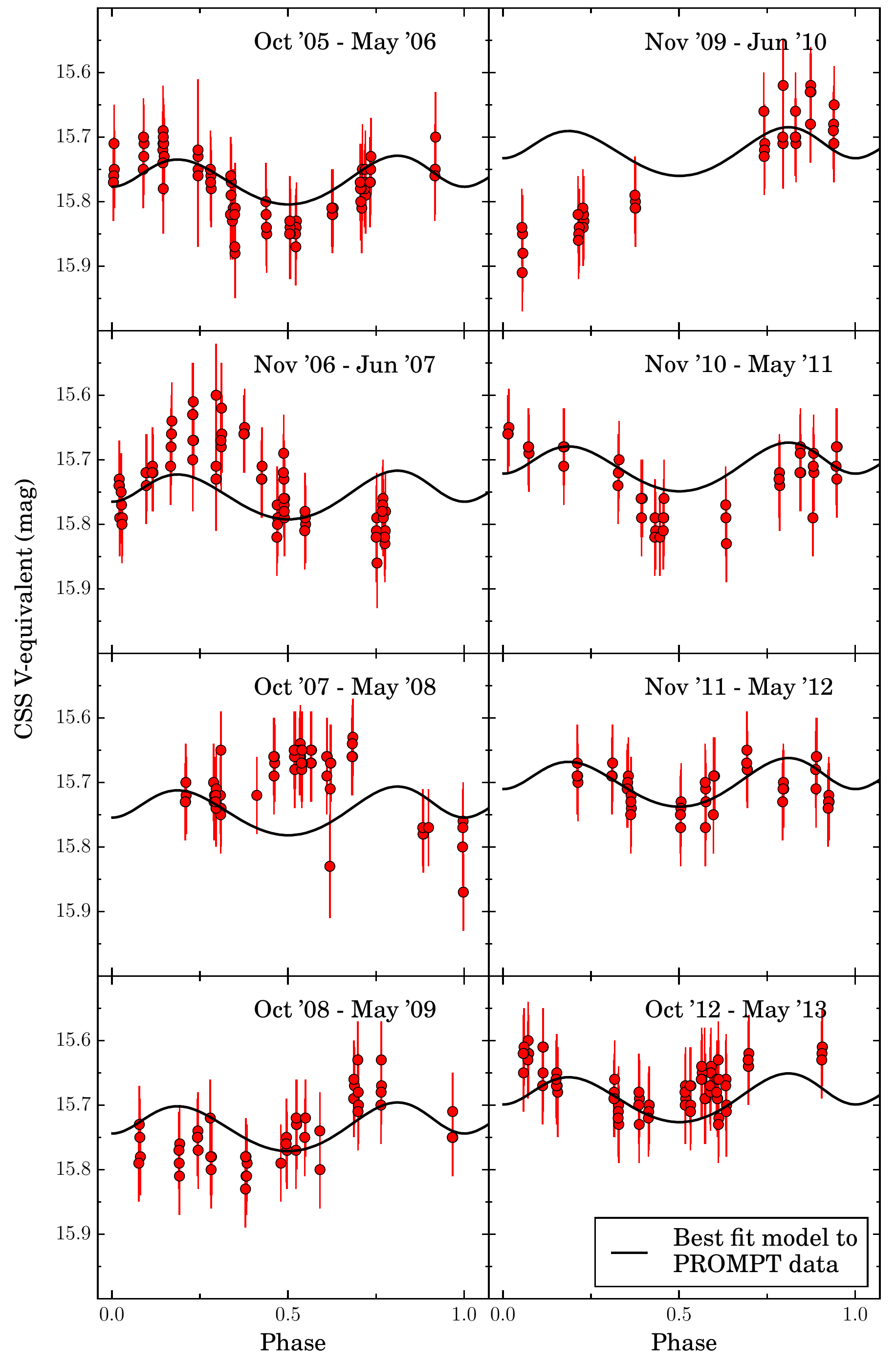}
    \centering
    \caption{The long-term CSS data split into eight separate epochs (first 4 on left, last 4 on right) and phased on the spectroscopic period. The best fit ELC model (with a disk) from \S~\ref{sec:inclination} is shown in black, adjusted for the linear trend seen in Figure~\ref{fig:catalina}. The shape of the light curve does not appear consistent with typical ellipsoidal variations across all epochs.}
    \label{fig:catalina_phased}
\end{figure}

To show the CSS photometry in more detail, we break the data into eight chunks corresponding roughly to the separate epochs visible in Figure~\ref{fig:catalina}. When we phase the data on the spectrocopic period, strange behavior is apparent (Figure~\ref{fig:catalina_phased}). The earliest and latest data show variations more consistent with the ellipsoidal variations found in the PROMPT data (black lines, adjusted for the linear brightening trend), though the phase coverage is not ideal. However, the epochs that bracket the decrease in gamma rays do not show the same variations, with large changes in either the phase of the variations, or no clear variations at all. For instance, the worst fit to the best ELC model occurs for the data taken between 2009 November and 2010 June ($\chi^2$/dof = 82.3/36), while the data from 2005 October to 2006 May clearly matches the model much better ($\chi^2$/dof = 23.2/71). This strange phenomenology as a function of epoch is similar to that observed by \citet{vanStaden16} for the optical light curves of the redback PSR J1723--2837, which they explain with distinct groups of starspots that vary in number and lifespan over time.

\section{Discussion and Conclusions}
\label{sec:conclusions}
Continuing our program to follow up on unassociated \emph{Fermi} sources using photometry and spectroscopy, we have shown that the previously unidentified $\gamma$-ray source 2FGL J0846.0+2820 is likely associated with a heavy neutron star of roughly 2~$M_{\odot}$ with a giant secondary companion ($M_2 \sim 0.8\,M_{\odot}$) in an 8.133 d orbit.

Unrelated variable stars will exist in the error ellipses of some unassociated \emph{Fermi} sources. For instance, the space density of CRTS variable stars is about 2.7 per deg$^2$ at the latitude of 2FGL J0846.0+2820. It is therefore not surprising that, in addition to the source that is the subject of this paper, there is another periodic optical variable (CSS J084632.3+282524) present inside the newly determined Pass 8 99\% confidence ellipse. This other source is classified as a contact binary with an optical period of 0.36 days \citep{Marsh17}. Based on the massive, invisible primary inferred for the original CSS source (CSS J084621.9+280839), and the rarity of compact binaries in the field, the evidence for an association between this source and the \emph{Fermi} $\gamma$-ray emission is strong.

An analogous example is the recent discovery of the likely redback MSP counterpart to the \emph{Fermi} source 3FGL J0838.8--2829. \citet{Halpern17} observed the 3FGL field in the optical and X-rays, motivated by the presence of a cataclysmic variable (CV) that showed variability in these bands. However, it was quickly determined that this CV was unlikely to be the $\gamma$-ray source. Instead, optical photometric and spectroscopic observations suggested a previously unknown X-ray/optical variable source, which showed properties consistent with redback MSPs in their pulsar state, was the correct counterpart to the $\gamma$-ray source \citep[see also,][]{Halpern17_2}.

Forthcoming X-ray observations will help confirm the connection between the \emph{Fermi} $\gamma$-ray source 2FGL J0846.0+2820 and the optical binary CSS J084621.9+280839.

Nonetheless, this work shows that such positional coincidence searches between optical variables and unassociated \emph{Fermi} sources can be productive and may provide a relatively ``low-cost'' method of identifying new compact binaries in the Galaxy.

Overall, 2FGL J0846.0+2820 is very similar to the recently discovered $\gamma$-ray bright binary 1FGL J1417.7--4407 \citep{Strader15}, which was independently found to host a radio MSP \citep{Camilo16}. This system has a heavy neutron star primary and giant secondary in a long 5.4 d orbit. The high primary mass and long period of 2FGL J0846.0+2820 are also similar to the famous MSP J1614--2230, though this system has a white dwarf secondary and has ceased mass transfer \citep{Demorest10}.

Similarly to the conclusions reached by \citet{Strader15} and \citet{Camilo16} for 1FGL J1417.7--4407, 2FGL J0846.0+2820 has properties consistent with a system on the standard evolutionary track of low-mass X-ray binaries that started Case B mass transfer after leaving the main sequence and whose orbital periods will increase over time \citep[e.g.,][]{Tauris99}. These systems do not fit into the existing classes of black widow or redback neutron star binaries, and we have suggested that an apt name for this new subclass is ``huntsman", after a large spider that does not engage in sexual cannibalism.

One relevant difference between 1FGL J1417.7--4407 and 2FGL J0846.0+2820 is the non-zero eccentricity of the latter. Nearly all binary MSPs in the field have circular orbits due to tidal interactions that occur when the system undergoes mass transfer, spinning up the pulsar. It is possible that 2FGL J0846.0+2820 has only partially completed the recycling process and not yet circularized, or that a non-zero eccentricity has emerged due to the interaction of the binary with a circumbinary disk, as has been theorized for some MSP systems \citep{Antoniadis14}.

It is now clear that both rotation-powered millisecond pulsars and accreting neutron stars can be associated with $\gamma$-ray emission. Such emission is generally weak for millisecond pulsars but is apparently ubiquitous in these systems, while it has been observed in only a small number of low-mass X-ray binaries. These latter systems all belong to the class of transitional millisecond pulsars or have other phenomenological similarities with this class. For two of the three systems that have shown actual transitions, the $\gamma$-ray flux is observed to be higher in the accreting state.

Thus, it would be natural to associate the decrease in $\gamma$-ray flux for 2FGL J0846.0+2820 with a transition from a disk to a pulsar state in mid-2009. As we discussed in \S~\ref{sec:longtermbrightness}, there are some unusual features in the optical light curve around that possible transition time. 

The issue with a direct analogy to the other transitional millisecond pulsars is that the higher variations in the optical flux were transient, with no long-term evidence of a state change. For example, the 2013 state change in PSR J1023+0038 was associated with an optical brightening of nearly 1 mag \citep{Bogdanov15}. Hence, these optical data are more consistent with a ``glitch" in the binary than a full-fledged state change. However, the relatively nearby main-sequence companion of PSR J1023+0038 is much less luminous than the more distant giant secondary in 2FGL J0846.0+2820. If the companion to PSR J1023+0038 were replaced with the giant in 2FGL J0846.0+2820, the secondary would swamp the disk, resulting in an inferred disk fraction that falls well below 10\%. Therefore, similar state changes in 2FGL J0846.0+2820 might not be associated with such obvious changes in the optical brightness.

Given that the mechanism for these transitions is not understood, and the notable physical differences between these hunstmen and the redback transitional millisecond pulsars, a difference in phenomenology of the optical and $\gamma$-ray emission may not be unexpected.

Some evidence for this can be found in 1FGL J1417.7--4407. In this system there is strong, double-peaked H$\alpha$ observed at nearly all epochs from early 2013 to the present (with the latter statement from continuing spectroscopic monitoring with SOAR). \citet{Strader15} took this as evidence for an accretion disk. This system also has a hard X-ray spectrum and a high ratio of $\gamma$-ray to X-ray flux, consistent with transitional MSPs in their disk states. However, this simple interpretation was challenged by the discovery of an MSP associated with this source by \citet{Camilo16}. The overlap between the observation epochs of these two studies show that the source was active as an MSP while it also appeared to have an accretion disk. It is possible that a disk is present but that the accreted material is not reaching the surface of the neutron star, perhaps due to ejection in a propeller \citep{Strader15}. Alternatively, the double-peaked H$\alpha$ could be due to complex line profiles associated with a wind \citep{Camilo16} or an intra-binary shock that forms due to the interaction of the pulsar wind outflows with the disk and/or material from the companion \citep[e.g.][]{Bogdanov11, Rivera17}.

At present the evidence for an accretion disk in 2FGL J0846.0+2820 is mixed. The gradual optical brightening and H$\alpha$ emission suggests a disk could be present. However, the current absence of gamma rays, the dominance of the photometric light curve by ellipsoidal variations, and our best-fit ELC models favor a low-mass disk at the most. Given that a millisecond pulsar was detected in 1FGL J1417.7--4407 \citep{Camilo16} despite the presence of strong double-peaked H$\alpha$ suggesting a more substantial accretion disk \citep{Strader15}, a search for a radio pulsar in 2FGL J0846.0+2820 is well-motivated.

The apparent variability in the \emph{Fermi} data along with the peculiar long-term optical light curve suggest we are far from fully understanding 2FGL J0846.0+2820. However, this work provides convincing evidence for the presence of a heavy neutron star primary with a giant secondary in a relatively wide orbit. These long-period $\gamma$-ray bright binaries with massive neutron stars and giant companions likely represent the late phases of \emph{typical} MSP binary formation in the Galactic field, phases which until recently had been unobserved. Further characterizations of similar systems would also provide new insight on the relationship between MSPs and low-mass X-ray binaries.

\section*{Acknowledgements}

We thank the anonymous referee for their helpful suggestions that improved the clarity of this paper. S.J.S. thanks Stephan Frank (Ohio State) for observing help provided during the on-site MDM run. Support from NASA grant NNX15AU83G is gratefully acknowledged. J.S. is supported by a Packard Fellowship. Based on observations obtained at the Southern Astrophysical Research (SOAR) telescope, which is a joint project of the Minist\'{e}rio da Ci\^{e}ncia, Tecnologia, e Inova\c{c}\~{a}o (MCTI) da Rep\'{u}blica Federativa do Brasil, the U.S. National Optical Astronomy Observatory (NOAO), the University of North Carolina at Chapel Hill (UNC), and Michigan State University (MSU). Some of the data presented herein were obtained at the W.M. Keck Observatory, which is operated as a scientific partnership among the California Institute of Technology, the University of California and the National Aeronautics and Space Administration. The Observatory was made possible by the generous financial support of the W.M. Keck Foundation. The \emph{Fermi} LAT Collaboration acknowledges generous ongoing support from a number of agencies and institutes that have supported both the development and the operation of the LAT as well as scientific data analysis. These include the National Aeronautics and Space Administration and the Department of Energy in the United States, the Commissariat \`{a} l'Energie Atomique and the Centre National de la Recherche Scientifique / Institut National de Physique Nucl\'{e}eaire et de Physique des Particules in France, the Agenzia Spaziale Italiana and the Istituto Nazionale di Fisica Nucleare in Italy, the Ministry of Education, Culture, Sports, Science and Technology (MEXT), High Energy Accelerator Research Organization (KEK) and Japan Aerospace Exploration Agency (JAXA) in Japan, and the K. A. Wallenberg Foundation, the Swedish Research Council and the Swedish National Space Board in Sweden. Additional support for science analysis during the operations phase is gratefully acknowledged from the Istituto Nazionale di Astrofisica in Italy and the Centre National d'\'{E}́tudes Spatiales in France. This work performed in part under DOE Contract DE-AC02-76SF00515. Portions of this research performed at NRL are sponsored by NASA DPR S-15633-Y.

\bibliography{report}

\end{document}